Second institution and/or address
This line break forced

# Realization of a Spin-1/2 Hexagonal-Plaquette Chain with Ising-Like Anisotropy

Hironori Yamaguchi[1,2], Shunsuke C. Furuya[3,4], Yu Tominaga[1], Koji Araki[5], and Masayuki Hagiwara[6]
[1]*Department of Physics, Osaka Metropolitan University, Osaka 558-8585, Japan*
[2]*Innovative Quantum Material Center (IQMC), Osaka Metropolitan University, Osaka 558-8585, Japan*
[3]*Department of Liberal Arts, Saitama Medical University, Moroyama, Saitama 350-0495, Japan*
[4]*Institute for Solid State Physics, the University of Tokyo, Chiba 277-8581, Japan*
[5]*Department of Applied Physics, National Defense Academy, Kanagawa 239-8686, Japan*
[6]*Center for Advanced High Magnetic Field Science (AHMF),
Graduate School of Science, Osaka University, Osaka 560-0043, Japan*
(Dated: July 17, 2025)

We present the realization of a spin-1/2 hexagonal-plaquette chain with Ising anisotropy, an unexplored quantum spin model that serves as a platform for investigating anisotropic quantum magnetism. Specific heat at zero field reveals a sharp peak at $T_{\rm N} = 1.0$ K, indicating a phase transition to a Néel order stabilized by interchain couplings. A perturbative analysis maps the system onto an effective spin-1/2 Ising-like chain, supporting the presence of an anisotropy-induced excitation gap. Furthermore, the interchain interactions may induce discrete excitations in the spinon continuum, reminiscent of Zeeman ladder physics observed in related 1D Ising-like systems. These results establish a well-defined model system for correlated spin phenomena in anisotropic magnets and highlight a route for engineering Ising-like quantum states in molecular-based frameworks.

PACS numbers: 75.10.Jm,

## I. INTRODUCTION

Quantum spin systems provide a fundamental platform for studying exotic quantum states and many-body interactions, playing a crucial role in quantum magnetism, spintronics, and quantum information science. Among these, one-dimensional (1D) Ising-like quantum magnets exhibit unique properties due to the interplay between anisotropic interactions and quantum fluctuations [1–3]. Their strong Ising anisotropy stabilizes ordered states while allowing for the emergence of unique excitations, including fractionalized spinon dynamics and quantum phase transitions. These systems serve as a crucial testing ground for understanding the effects of anisotropic quantum correlations.

Experimentally, Co-based materials such as CsCoX$_3$ (X = Cl, Br) [4–8] and ACo$_2$V$_2$O$_8$ (A = Ba, Sr) [9–12] have served as prototypical systems for realizing 1D spin-1/2 Ising-like antiferromagnets. These materials exhibit strong exchange anisotropy originating from the spin-orbit coupling of Co$^{2+}$ ions in their crystal field environments. Experimental studies have revealed that in such systems, the spinon continuum–typically expected in 1D spin-1/2 chains–can become discretized due to weak interchain interactions or external perturbations. This phenomenon, known as Zeeman ladder discretization [13], manifests as quantized excitations and has been both theoretically predicted and experimentally verified in these Co-based compounds. Among them, BaCo$_2$V$_2$O$_8$ has garnered particular attention as an ideal platform for studying quantum phenomena associated with Ising anisotropy. Not only does it exhibit Zeeman ladder excitations arising from interchain interactions [14], but it also undergoes an order-to-disorder quantum phase transition when a magnetic field is applied along the easy axis [15–18]. Moreover, recent studies have revealed that applying transverse magnetic fields induces additional exotic quantum phases, further enriching the physics of these systems[19–22]. These findings underscore the importance of 1D Ising-like chains in revealing fundamental aspects of quantum magnetism. However, despite these advances, the range of available Ising-like quantum spin systems remains limited, and tunable model materials have been scarce due to the structural constraints of inorganic compounds.

In this paper, we report the realization of a spin-1/2 hexagonal-plaquette chain with strong Ising anisotropy, an unexplored 1D spin model. This system provides a platform for investigating anisotropic quantum magnetism in a geometrically distinct framework. We synthesized a organic-based Co complex that realizes this spin model. Specific heat measurements reveal a phase transition to a Néel-ordered state at $T_{\rm N} = 1.0$ K. Below $T_{\rm N}$, we observe an anomalous Schottky-like shoulder, suggesting the formation of an excitation gap due to anisotropic spin correlations. Perturbative analysis maps the system onto an effective spin-1/2 Ising-like chain, supporting the presence of a gapped excitation spectrum. The interchain interactions may influence the excitation spectrum in a way that resembles Zeeman ladder-like discretization, as reported in other 1D Ising-like systems.

## II. EXPERIMENTAL

We synthesized $p$-Py-V via the conventional procedure for producing the verdazyl radical [23]. A solution of $Co(NO_3)_2 \cdot 6H_2O$ (87.3 mg, 0.30 mmol) in 2 ml ethanol was slowly added to a solution of $p$-Py-V (188.6 mg, 0.60 mmol) in 6 ml of $CH_2Cl_2$ and stirred for 30 min. A dark-brown crystalline solid of $(p\text{-Py-V})_2[Co(NO_3)_2]$ was separated by filtration. Single crystals were obtained via recrystallization using $CHCl_3$ at 40°C.

The X-ray intensity data were collected using a Rigaku XtaLAB Synergy-S instrument. Magnetization measurements were conducted using a commercial superconducting quantum interference device (MPMS, Quantum Design). Specific heat measurements were performed using a commercial calorimeter (PPMS, Quantum Design). X-band electron spin resonance (ESR) measurements were conducted using a JEOL spectrometer. All experiments were performed using powder samples. For the specific heat and ESR measurements, samples were fixed with grease to suppress the orientation in the external field direction. Molecular orbital (MO) calculations were performed using the unrestricted Becke three-parameter Lee-Yang-Parr functional (UB3LYP) method with a basis set of 6-31G and 6-31G(d, p) [24]. All calculations were performed using the GAUSSIAN09 software package.

## III. RESULTS AND DISCUSSION

### A. Crystal structure and spin model

The crystallographic parameters of $(p\text{-Py-V})_2[Co(NO_3)_2]$ are provided in Table I. Figure 1(a) shows the molecular structure, with the Co atom surrounded by two radicals and two nitrate ligands, creating a six-coordinate environment. Table II lists the bond lengths and angles of the Co atoms. The verdazyl radical $p$-Py-V and $Co^{2+}$ have spins of 1/2 and 3/2, respectively. As a twofold rotational axis parallel to the $b$ axis runs through the position of the Co atom, two $p$-Py-V radicals in the molecule are crystallographically equivalent. In the $p$-Py-V, approximately 61% of the total spin density is distributed in the central ring consisting of four nitrogen atoms, leading to a localized spin system [25, 26]. The MO calculations were used to evaluate exchange interactions. The intramolecular interaction was ferromagnetic (F) and evaluated as $J_1/k_B$ = -31 K. Because MO calculations tend to overestimate the intramolecular interactions between verdazyl radicals and transition metals, it is anticipated that the actual value of $J_1$ will be approximately half of the MO evaluation [27, 28]. For intermolecular interactions between radical spins, we found a predominant antiferromagnetic (AF) interaction ($J_2/k_B$ = 2 K), forming a spin-1/2 1D uniform chain along the $c$ axis, as shown in Fig. 1(b). Each radical spin ($\sigma$) in the 1D spin

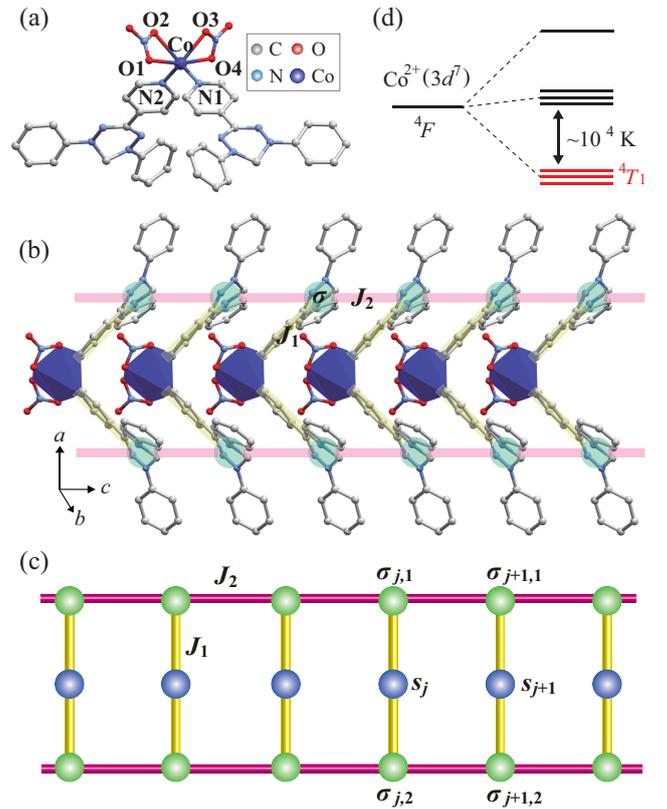

FIG. 1: (a) Molecular structure of $(p\text{-Py-V})_2[Co(NO_3)_2]$, leading to intramolecular exchange interaction $J_1$ between the radical and Co spins. (b) Crystal structure forming the hexagonal-plaquette chain along the $c$ axis. Hydrogen atoms are excluded to enhance clarity. The green nodes represent the spin-1/2 of the radicals. The thick lines represent exchange interactions. (c) Spin-1/2 hexagonal-plaquette chain comprising $J_1$ and $J_2$. $\sigma$ and $s$ denote the spins on the radical and $Co^{2+}$, respectively. (d) Electronic configuration of high-spin $Co^2$ with 3d electrons leading to octahedral ligand field splitting, resulting in the formation of the lowest orbital triplet $^4T_1$.

chain is coupled with the Co spin ($s$) via $J_1$, yielding a hexagonal-plaquette chain, as shown in Fig. 1(b) and 1(c). Furthermore, the molecular arrangement exhibits no significant MO overlap between the 1D structures, enhancing the 1D nature of the present spin lattice. For the high-spin $Co^{2+}$, the ground state in an octahedral environment is the orbital triplet $^4T_1$, as shown in Fig. 1(d). The introduction of spin-orbit coupling and crystal field distortion causes the split of $^4T_1$ into six Kramers doublets [29, 30], yielding a effective spin-1/2 ($s$) with anisotropic $g$ values and exchange interactions at low temperatures. Then, the spin Hamiltonian of the hexagonal-plaquette chain composed of $\sigma$ and $s$ is given

TABLE I: Crystallographic data of $(p\text{-Py-V})_2[\text{Co}(\text{NO}_3)_2]$.

| Formula | $\text{C}_{38}\text{H}_{32}\text{CoN}_{12}\text{O}_6$ |
| --- | --- |
| Crystal system | Orthorhombic |
| Space group | $P2_12_12$ |
| Temperature (K) | 100(2) |
| $a$ (Å) | 17.8235(12) |
| $b$ (Å) | 18.7872(14) |
| $c$ (Å) | 5.3707(4) |
| $V$ (Å$^3$) | 1798.4(2) |
| $Z$ | 2 |
| $D_{\text{calc}}$ (g cm$^{-3}$) | 1.499 |
| Total reflections | 1958 |
| Reflection used | 1835 |
| Parameters refined | 258 |
| $R$ $[I > 2\sigma(I)]$ | 0.0510 |
| $R_w$ $[I > 2\sigma(I)]$ | 0.1294 |
| Goodness of fit | 1.058 |
| CCDC | 2429446 |

TABLE II: Bond lengths (Å) and angles (°) related to the Co atom in $(p\text{-Py-V})_2[\text{Co}(\text{NO}_3)_2]$.

| Co–N1 | 2.04 | N1–Co–O1 | 89.5 |
| --- | --- | --- | --- |
| Co–N2 | 2.04 | O1–Co–O2 | 52.6 |
| Co–O1 | 2.24 | O2–Co–O4 | 116.9 |
| Co–O2 | 2.23 | O4–Co–N1 | 96.6 |
| Co–O3 | 2.23 | N1–Co–N2 | 115.5 |
| Co–O4 | 2.24 | N2–Co–O2 | 98.4 |
| | | O2–Co–O3 | 79.5 |
| | | O3–Co–N1 | 98.4 |
| | | N2–Co–O1 | 96.6 |
| | | O1–Co–O3 | 116.9 |
| | | O3–Co–O4 | 52.6 |
| | | O4–Co–N2 | 89.5 |

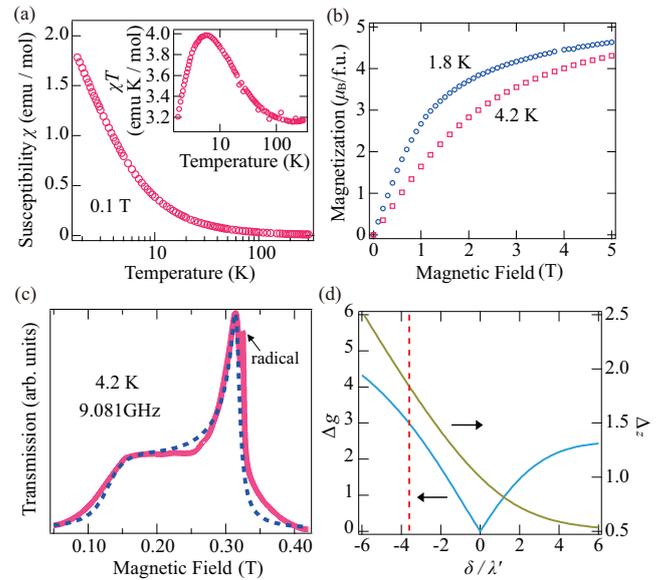

FIG. 2: (a) Temperature dependence of magnetic susceptibility ($\chi = M/H$) of $(p\text{-Py-V})_2[\text{Co}(\text{NO}_3)_2]$ at 0.1 T. The inset shows corresponding $\chi T$ values. (b) Magnetization curves of $(p\text{-Py-V})_2[\text{Co}(\text{NO}_3)_2]$ at 1.8 and 4.2 K. (c) ESR absorption spectrum of $(p\text{-Py-V})_2[\text{Co}(\text{NO}_3)_2]$ at 9.081 GHz and 4.2 K. The broken line represents the powder pattern simulation. The sharp resonance signal at approximately 0.33 T is attributed to spins on the radical. (d) Anisotropic difference in the $g$ values for the ground Kramers doublet and expected anisotropy of the effective exchange interaction between the spins on the radical and Co$^{2+}$ ion, i.e., $J_1$, as a function of $\delta_z/\lambda'$ with $k = 0.6$. The vertical line shows $\delta_z/\lambda' = -3.6$ for the present compound.

by

$$\mathcal{H} = J_1 \sum_j \sum_{\alpha=1,2} (s_j^x \sigma_{j,\alpha}^x + s_j^y \sigma_{j,\alpha}^y + \Delta_z s_j^z \sigma_{j,\alpha}^z)$$
$$+ J_2 \sum_j \sum_{\alpha=1,2} \boldsymbol{\sigma}_{j,\alpha} \cdot \boldsymbol{\sigma}_{j+1,\alpha} - g\mu_B \sum_j \boldsymbol{H} \cdot \boldsymbol{\sigma}_j - \mu_B \sum_j \boldsymbol{H} \tilde{\mathbf{g}} \boldsymbol{s}_j,$$
(1)

where $\Delta_z$ denotes Ising-like anisotropy, $\boldsymbol{H}$ denotes the external magnetic field, $g$ denotes the $g$ of the radical spin, and $\tilde{\mathbf{g}}$ denotes the $g$-tensor of the Co spin. The diagonal components for the principal axes of the $g$-tensor are $g_x$, $g_y$, and $g_z$ and the other components are zero.

### B. Magnetization

The magnetic susceptibility ($\chi = M/H$) at 0.1 T is shown in Fig. 2(a). The temperature dependence of $\chi T$ increases with decreasing temperature until approximately 6 K, which demonstrates the dominant contributions of F $J_1$. The contributions of AF $J_2$ are indicated by the decrease of $\chi T$ in the lower temperature region. Figure 2(b) displays the magnetization curve at 1.8 K and 4.2 K. These results are reminiscent of paramagnetic-like behavior, which is due to the small energy scale of AF $J_2$ forming the spin chain. Considering the typical values of the Van Vleck paramagnetism, 0.01-0.02 $\mu_B$/Co$^{2+}$, in Co-based compounds [31, 32], the magnetic moment approaches to $\sim$4.5 $\mu_B$ in the high-field region. In Co-based systems with strong easy-axis anisotropy, it is well known that unfixed powder samples tend to align along the direction of an external magnetic field. This behavior has also been observed in our previous studies on Co-based complexes [33–35]. Such field-induced alignment allows measurements to reflect the response along the easy axis. Accordingly, the asymptotic value indicates the full polarization of both $\boldsymbol{\sigma}$ with $g\sim$2.0 and $\boldsymbol{s}$ with $g_z\sim$5.0 for the easy axis, i.e., $2g\sigma+g_zs \approx 4.5$. This interpretation is also supported by the anisotropic $g$ values extracted from the following ESR results.

## C. Electron spin resonance

Figure 2(c) shows the ESR spectrum at 9.081 GHz and at 4.2 K, where the paramagnetic resonance is expected to be observed. As the powder sample used in this measurement is restrained from orienting in the external field direction by mixing it with grease, the observed signal corresponds to the powder pattern. A sharp resonance signal characteristic of organic radical systems is observed at 0.33 T, confirming the isotropic nature of the verdazyl radical with $g=2.0$. The other resonance signals extended to the wide field range originate from the Co spin and reflect the anisotropic $g$ values. We simulated the ESR spectrum assuming paramagnetic resonance with anisotropic $g$ values [33–35], leading to the agreement with the experimental data using $g_x = g_y = 2.05$ and $g_z = 5.05$, as shown in Fig. 2(c).

We examined the evaluated magnetic anisotropy of $Co^{2+}$ in the octahedral crystal field. The total orbital and spin angular momenta of the high-spin $Co^{2+}$ ion are $L = 3$ and $S = 3/2$, respectively. The ground orbital triplet $^4T_1$ has an effective orbital angular momentum $l = 1$, and the orbital degeneracy is removed by the axial crystal fields, $\delta_z$, leading to axial magnetic anisotropy. The Hamiltonian, derived from perturbative to spin-orbit coupling and crystal fields, is expressed as follows:

$$\mathcal{H} = \frac{3}{2}\lambda' \boldsymbol{l} \cdot \boldsymbol{S} - \delta(l_z^2 - \frac{2}{3}), \quad (2)$$

where $\lambda' = k\lambda$, $\lambda$ denotes the spin-orbit coupling constant, and $k$ denotes the orbital reduction factor originating from the admixture between the $3d$ electron and $p$ electrons in the ligands [29, 30]. Here, we constructed a $12 \times 12$ Hamiltonian matrix using the basis states $|l_z, S_z\rangle$. The energy spectrum was obtained by numerically solving the secular equation as a function of $\delta z/\lambda'$, with all other parameters held constant. A sizable energy gap of several hundred kelvin separates the ground-state Kramers doublet from the first excited state, allowing the low-temperature magnetic properties to be effectively described within the subspace of the doublet. Under an applied magnetic field, this lowest Kramers doublet $\psi\pm$ splits due to the Zeeman interaction defined by

$$\mathcal{H}_Z = \mu_B(-\frac{3}{2}k\boldsymbol{l} + 2\boldsymbol{S})\cdot \boldsymbol{H}. \quad (3)$$

Considering the energy shift of Kramers doublet as the Zeeman splitting of the effective spin-1/2, i.e., $\boldsymbol{s}$, the $g$ values for the principal axes are given by

$$g_z = \pm 2\langle\psi_\pm|(-\frac{3}{2}kl_z + 2S_z)|\psi_\pm\rangle, \quad (4)$$

$$g_x = \langle\psi_\pm|[-\frac{3}{2}k(l^+ + l^-) + 2(S^+ + S^-)]|\psi_\mp\rangle, \quad (5)$$

$$g_y = \mp\langle\psi_\pm|[-\frac{3}{2}k(l^- - l^+) + 2(S^- - S^+)]|\psi_\mp\rangle, \quad (6)$$

where $l^\pm = l_x \pm i l_y$ and $S^\pm = S_x \pm i S_y$. Consequently, the anisotropic difference in the $g$ values evaluated from the ESR analysis is explained by the parameters $\delta/\lambda' = -3.6$ assuming $k = 0.6$, which is close the typical value for the verdazyl-based Co complexes [33–35], as shown in Fig. 2(d). The anisotropy of the exchange interaction between $\boldsymbol{s}$ and $\boldsymbol{\sigma}$, $\Delta_z$, is also evaluated as a function of $\delta/\lambda'$. Considering the matrix elements of $S$ written as

$$\langle\psi_\pm|S_z|\psi_\pm\rangle = \pm\frac{p}{2}, \langle\psi_\pm|S_x|\psi_\mp\rangle = \frac{q_x}{2}, \langle\psi_\pm|S_y|\psi_\mp\rangle = \mp i\frac{q_y}{2}, \quad (7)$$

the true spin-3/2 on $Co^{2+}$ is replaced by

$$S_z = ps^z, S_x = q_x s^x, S_x = q_y s^y. \quad (8)$$

Then, we obtained $\Delta_z = p/q_x$ ($=p/q_y$) as a function of $\delta_z/\lambda'$, indicating that the complex exhibits Ising-type anisotropy with $\Delta_z = 1.8$, as shown in Fig. 2(d).

## D. Specific heat

Figure 3(a) shows the experimental result for the specific heat $C_p$ at zero field. The magnetic contributions are expected to be dominant in the low-temperature regions considered here. We found a sharp peak at $T_N = 1.0$ K, demonstrating a phase transition to an AF ordered state induced by interchain couplings. Although the value of $T_N$ may seem relatively high compared to the energy scale of the AF correlations expected from the $\chi T$ behavior, a Néel state, which is the expected ground state of the present system as discussed later, can induce long-range order due to weak but finite interchain interactions. This provides a natural explanation for the relatively high $T_N$ despite the predominantly 1D nature of the magnetic interactions. The rounded behavior at approximately 2 K is considered to originate from the 1D sort-range correlations in the hexagonal-plaquette chain, as shown in Fig. 3(b). Note that an anomalous shoulder was observed below $T_N$, which becomes more pronounced when plotted as $C_p/T$, as shown in Fig. 3(a). This behavior suggests the presence of an energy gap $\Delta$ even in the ordered phase. Assuming that the low-temperature specific heat follows the form expected for 1D gapped systems, $C_p \propto T^{-2} \exp(-\Delta/k_B T)$ [36, 37], we estimated the gap as $\Delta = 1.7$ K. Since entropy is obtained via integration of $C_p/T$, the contribution from the low-temperature region is essential for accurate evaluation. However, due to the limited experimental temperature range, we extrapolated the data using the gapped behavior fit. The resulting entropy shift from this extrapolation was relatively small (0.37 J/mol.K), indicating that the overall estimate is not significantly affected, as shown in Fig. 3(c). The entropy change below $T_N$ reaches approximately two-thirds of the full magnetic entropy for a spin-1/2 system, $R\ln 2 \approx 5.76$ J/mol.K. This indicates that significant short-range order associated with the effective 1D spin-1/2 model–discussed in the following section–is already well developed above $T_\text{N}$. By applying magnetic fields, the phase transition temperature shifts to the lower temperature region and broadens reflecting the anisotropy effect inherent to the powder sample, as shown in Fig. 3(d).

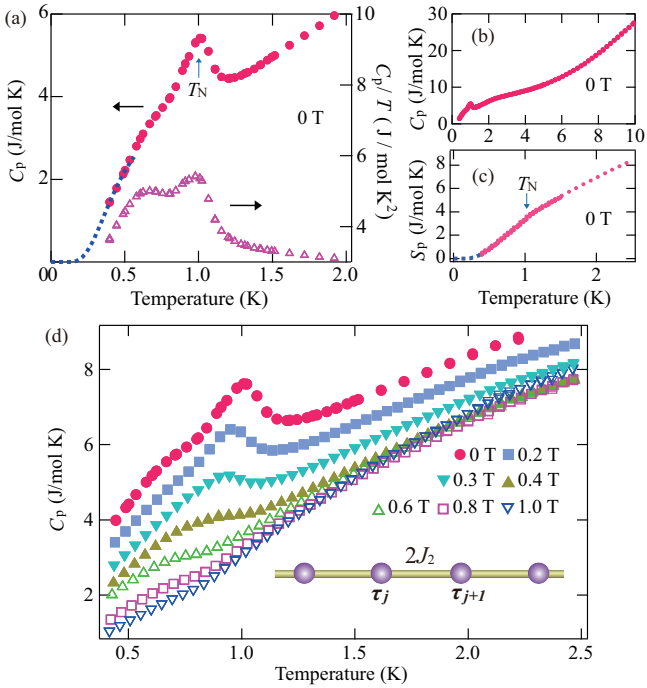

FIG. 3: (a) Temperature dependence of the specific heat $C_\text{p}$ and $C_\text{p}/T$ of $(p\text{-Py-V})_2[\text{Co}(\text{NO}_3)_2]$ at zero field. The broken line indicates the fitting curve with the low-temperature specific heat for 1D gapped systems. The arrows indicate the phase-transition temperature $T_\text{N}$. (b) Temperature dependence of $C_\text{p}$ at zero field with the expanded temperature regime. (c) Temperature dependence of the entropy $S_\text{p}$ at zero field. The experimental values of $S_\text{p}$ have been shifted up to overlap with the values evaluated from the fitting curve of $C_\text{p}$. (d) Temperature dependence of $C_\text{p}$ at various magnetic fields. For clarity, the values for 0, 0.2, 0.3, 0.4, and 0.6 T have been shifted up by 2.2, 1.6, 1.0, 0.5, and 0.3 J/ mol K, respectively. The illustration depicts the effective Ising-like chain formed by the effective spin-1/2 degree of freedom, $\boldsymbol{\tau}$, in the low-temperature regime for $|J_1| \gg J_2$ and $\Delta_z \gg 1$.

### E. Low-temperature effective spin model

We discuss the ground state of the hexagonal plaquette chain described by Eq. (1) for $|J_1| \gg J_2$ and strong Ising anisotropy $\Delta_z \gg 1$. Since the local eigenstates are primarily determined by $J_1$, we do not explicitly consider the difference in $g$ values between $s_j$ and $\sigma_{j,\alpha}$ in this analysis, leading to the use of the normalized field $h$. In the present regime, the system can be regarded as a collection of nearly independent trimers composed of $s_j$ and neighboring spins $\sigma_{j,1}$ and $\sigma_{j,2}$, where the strong exchange interaction $J_1$ dominates the local spin configurations. The inter-trimer interactions, mediated by the weaker coupling $J_2$, introduce perturbative corrections. To systematically construct the effective model, we first analyze the local eigenstates of an isolated trimer, where $s_j$ interacts with both $\sigma_{j,1}$ and $\sigma_{j,2}$ through the dominant interaction $J_1$. The Hamiltonian for the trimer is given by

$$H_j = J_1 \sum_{\alpha=1,2} \left( s_j^x \sigma_{j,\alpha}^x + s_j^y \sigma_{j,\alpha}^y + \Delta_z s_j^z \sigma_{j,\alpha}^z \right) \\ - h \left( \sigma_{j,1}^z + \sigma_{j,2}^z + s_j^z \right). \quad (9)$$

Defining the composite spin $\boldsymbol{S}_\text{V} = \boldsymbol{\sigma}_{j,1} + \boldsymbol{\sigma}_{j,2}$, we classify the local spin states into total spin configurations, where $S_\text{V} = 1$ corresponds to triplet states and $S_\text{V} = 0$ to a singlet state. The eigenstates of $H_j$ can be labeled by three quantum numbers: $s_j^z = \pm 1/2$, $S_\text{V}^z = \pm 1, 0$, and the total spin $S_\text{V} = 1, 0$. For large Ising anisotropy $\Delta_z \gg 1$, consideration of the energy separation between the ground states and the first excited states, which is approximately $0.34|J_1|$ for $\Delta_z = 1.8$, leads to effective spin-1/2 degrees of freedom $\tau_j$, arising from the lowest-energy states $\left|\frac{3}{2}\right\rangle$ and $\left|-\frac{3}{2}\right\rangle$. Since the system has strong Ising anisotropy, the lowest-energy states are expected to be dominated by configurations with spins aligned along the $z$-axis. However, due to the presence of transverse exchange terms in $H_j$, quantum mixing occurs between other nearby spin configurations. The degree of this mixing is characterized by a mixing angle $\theta$, which quantifies the admixture of these configurations and controls the transverse components in the effective Hamiltonian. The mixing angle is given by

$$\tan\theta = \frac{\sqrt{8}}{\Delta_z + \sqrt{\Delta_z^2 + 8}}. \quad (10)$$

With the low-energy subspace identified, we incorporate the effect of the inter-trimer coupling $J_2$, which introduces interactions between neighboring effective spins. The full Hamiltonian can be written as $H = H_0 + V$, where $H_0 = \sum_j H_j$ describes the decoupled trimers, and $V = J_2 \sum_j \sum_{\alpha=1,2} \boldsymbol{S}_{j,\alpha} \cdot \boldsymbol{S}_{j+1,\alpha}$ accounts for the inter-trimer exchange. We treat $V$ perturbatively up to third order in $J_2$ by projecting the full Hamiltonian onto the low-energy subspace. At second order, the dominant contribution is an effective Ising interaction between neighboring effective spins, leading to the term:

$$H_\text{eff}^{(2)} = 2J_2 \sum_j \tau_j^z \tau_{j+1}^z - \frac{h}{3} \sum_j \tau_j^z. \quad (11)$$

At third order, virtual transitions between trimer states generate an additional transverse exchange term, resulting in an effective XXZ-type interaction:

$$H_\text{eff}^{(3)} = 2J_2 \sum_j \left[ \epsilon(\tau_j^x \tau_{j+1}^x + \tau_j^y \tau_{j+1}^y) + \tau_j^z \tau_{j+1}^z \right] - \frac{h}{3} \sum_j \tau_j^z. \quad (12)$$

Here, the anisotropy parameter $\epsilon$ reflects the strength of the transverse exchange interaction relative to the dominant Ising term and originates from third-order virtual processes involving intermediate trimer excitations. For large Ising anisotropy $\Delta_z \gg 1$, the parameter $\epsilon$ is then given by:

$$\epsilon = \frac{1}{8\sqrt{2}} \frac{\cos^3\theta \sin\theta}{(\Delta_z - \sqrt{2}\tan\theta)^2} \approx \frac{1}{8\Delta_z^3} \left(\frac{J_2}{J_1}\right)^2. \quad (13)$$

The obtained expression for $\epsilon$ explicitly demonstrates that the transverse components of the effective interaction originate from third-order virtual processes involving the original anisotropic exchange $J_1$ and the perturbative coupling $J_2$. As a result, the effective Hamiltonian acquires an XXZ form, where the anisotropy $\epsilon$ directly reflects the Ising-like nature of the original interactions. This establishes a clear mapping of the system onto a spin-1/2 Ising-like chain. The ground state of the effective model is expected to be a Néel state, which aligns well with experimental observations. Given this, the lowest excitation is naturally considered to be a gapped spinon continuum. Furthermore, specific heat measurements reveal an anomalous Schottky-like behavior that appears as a broad shoulder rather than a distinct peak. This feature suggests the formation of an excitation gap below $T_N$, consistent with the presence of anisotropy-induced spin correlations.

In our effective model, the excitation gap is given by $2J_2(1-2\epsilon)$ [38], and using the experimental gap of 1.7 K, we estimate $J_2/k_B \approx 1$ K, in agreement with the molecular orbital evaluation. In quasi-1D magnets, weak but finite interchain interactions not only stabilize long-range order but can also modify the excitation spectrum. In particular, they may discretize the spinon continuum into multiple gapped levels. This scenario is reminiscent of Zeeman ladder excitations, which have been reported in other Ising-like spin chains with weak interchain couplings [13, 14]. To estimate the strength of the interchain interaction $J'$, we use the formula [39]: $J'/k_B \simeq T_N/(1.28\sqrt{\ln(5.8J/(k_B T_N))})$, where $J$ is the dominant intrachain interaction, taken here as $2J_2$. This relation is derived under the assumption of well-developed intrachain correlations and a mean-field treatment of interchain coupling [39]. Therefore, the nature of intrachain correlations–whether Heisenberg-like or Ising-like– does not affect the validity of this relation, as long as the intrachain spin correlations are sufficiently developed. With $J_2/k_B \approx 1$ K and $T_N = 1.0$ K, we obtain $J'/k_B \approx 0.5$ K. This value should not be interpreted as a precise parameter but rather as an indicative scale that suggests the potential influence of interchain effects on the excitation spectrum. While our data do not definitively confirm the Zeeman ladder picture, the observed broad anomaly in the specific heat and the estimated interchain coupling suggest that such discretized excitations may indeed be present in this system.

## IV. SUMMARY

In summary, we synthesized a Co-based complex ($p$-Py-V)$_2$[Co(NO$_3$)$_2$], which realizes a spin-1/2 hexagonal-plaquette chain. ESR and spin-orbit coupling analysis confirmed anisotropic $g$ values and Ising-type interactions. Specific heat measurements indicated a phase transition to a Néel-ordered state driven by weak interchain couplings and a Schottky-like behavior in the lower temperature region. A perturbative analysis mapped the system onto an effective spin-1/2 1D Ising-like chain, supporting the presence of an anisotropy-induced excitation gap. Furthermore, the interchain effects may modify the excitation spectrum to produce discrete levels, potentially reminiscent of Zeeman ladder excitations observed in related Ising-like systems. This work not only expands the available library of spin systems but also provides a promising avenue for investigating anisotropic quantum states and their associated excitations. Beyond fundamental physics, the ability to engineer quantum spin systems has broader implications for quantum information processing, spintronics, and the exploration of exotic correlated phases.

### Acknowledgments

We thank C. Pichon for valuable discussions. The ESR measurement (or analysis) were performed at the Analytical Instrument Facility, Graduate School of Science, Osaka University. A part of this work was performed under the interuniversity cooperative research program of the joint-research program of ISSP, the University of Tokyo.

[1] F. D. M. Haldane, General relation of correlation exponents and spectral properties of one-dimensional Fermi systems: application to the anisotropic $S=1/2$ Heisenberg chain, Phys. Rev. Lett. **45**, 1358 (1980).

[2] C. N. Yang and C. P. Yang, One-dimensional chain of anisotropic spin-spin interactions. III. Applications, Phys. Rev. **151**, 258 (1966).

[3] P. Pfeuty, The one-dimensional Ising model with a transverse field, Ann. Phys. **57**, 79 (1970).

[4] W. P. Lehmann, W. Breitling, and R. Weber, Raman scattering study of spin dynamics in the quasi-1D Ising antiferromagnets CsCoCl$_3$ and CsCoBr$_3$, J. Phys. C **14**, 4655 (1981).

[5] S. E. Nagler, W. J. L. Buyers, R. L. Armstrong, and B. Briat, ,Ising-like spin-1/2 quasi-one-dimensional antiferromagnets: Spin-wave response in CsCoX$_3$ salts, Phys. Rev. B **27**, 1784 (1983).

[6] K. Amaya, H. Hori, I. Shiozaki, M. Date, M. Ishizuka,

T. Sakakibara, T. Goto, N. Miura, H. Kikuchi, and Y. Ajiro, High field magnetization of quasi one dimensional Ising antiferromagnet $CsCoCl_3$, J. Phys. Soc. Jpn. **59**, 1810 (1990).

[7] H. Shiba, Y. Ueda, K. Okunishi, S. Kimura, and K. Kindo, Exchange interaction via crystal-field excited states and its importance in $CsCoCl_3$, J. Phys. Soc. Jpn. **72**, 2326 (2003).

[8] S. Kimura, H. Onishi, K. Okunishi, M. Akaki, Y. Narumi, M. Hagiwara, K. Kindo, and H. Kikuchi, Magnetic excitation in the $S = 1/2$ Ising-like antiferromagnetic chain $CsCoCl_3$ in longitudinal magnetic fields studied by high-field ESR measurements, J. Phys. Soc. Jpn. **90**, 094701 (2023).

[9] R. Wichmann and Hk. Müller-Buschbaum, Neue verbindungen mit $SrNi_2V_2O_8$-Struktur: $BaCo_2V_2O_8$ und $BaMg_2V_2O_8$, Z. Anorg. Allg. Chem. **534**, 153 (1986).

[10] D. Osterloh and Hk. Müller-Buschbaum, Zur Kenntnis von $SrCo_2V_2O_8$ und $SrCo_2(AsO_4)_2$, Z. Naturforsch. B **49**, 923 (1994).

[11] Z. Wang, M. Schmidt, A. K. Bera, A. T. M. N. Islam, B. Lake, A. Loidl, and J. Deisenhofer, Spinon confinement in the one-dimensional Ising-like antiferromagnet $SrCo_2V_2O_8$, Phys. Rev. B **91**, 140404(R) (2015).

[12] A. K. Bera, B. Lake, F. H. L. Essler, L. Vanderstraeten, C. Hubig, U. Schollwöck, A. T. M. N. Islam, A. Schneidewind, and D. L. QuinteroCastro, Spinon confinement in a quasi-one-dimensional anisotropic Heisenberg magnet, Phys. Rev. B **96**, 054423 (2017).

[13] H. Shiba, Quantization of magnetic excitation continuum due to lnterchain coupling in nearly one-dimensional Ising-like antiferromagnets, Prog. Theor. Phys. **64**, 466 (1980).

[14] B. Grenier, S. Petit, V. Simonet, E. Canévet, L.-P. Regnault, S. Raymond, B. Canals, C. Berthier, and P. Lejay, Longitudinal and transverse zeeman Ladders in the Ising-like chain antiferromagnet, Phys. Rev. Lett. **114**, 017201 (2015).

[15] Z. He, T. Taniyama, T. Kyômen, and M. Itoh, Field-induced order-disorder transition in the quasi-one-dimensional anisotropic antiferromagnet $BaCo_2V_2O_8$, Phys. Rev. B **72**, 172403 (2005).

[16] S. Kimura, H. Yashiro, K. Okunishi, M. Hagiwara, Z. He, K. Kindo, T.Taniyama, and M. Itoh, Field-induced order-disorder transition in antiferromagnetic $BaCo_2V_2O_8$ driven by a softening of spinon excitation, Phys. Rev. Lett. **99**, 087602 (2007).

[17] S. Kimura, T. Takeuchi, K. Okunishi, M. Hagiwara, Z. He, K. Kindo,T. Taniyama, and M. Itoh, Novel ordering of an $S =1/2$ quasi-1d Ising-like antiferromagnet in magnetic field, Phys. Rev. Lett. **100**, 057202 (2008).

[18] S. Kimura, M. Matsuda, T. Masuda, S. Hondo, K. Kaneko, N. Metoki,M. Hagiwara, T. Takeuchi, K. Okunishi, Z. He, K. Kindo, T.Taniyama, and M. Itoh, Longitudinal spin density wave order in a quasi-1D Ising-like quantum antiferromagnet, Phys. Rev. Lett. **101**, 207201 (2008).

[19] Q. Faure, S. Takayoshi, S. Petit, V. Simonet, S. Raymond, L.-P. Regnault, M. Boehm, J. S. White, M. Månsson, C. Rüegg, P. Lejay, B.Canals, T. Lorenz, S. C. Furuya, T. Giamarchi, and B. Grenier, Topological quantum phase transition in the Ising-like antiferromagnetic spin chain $BaCo_2V_2O_8$, Nat. Phys. **14**, 716 (2018).

[20] Y. Cui, H. Zou, N. Xi, Z. He, Y. X. Yang, L. Shu, G. H. Zhang, Z. Hu, T. Chen, R. Yu, J. Wu, and W. Yu, Quantum criticality of the Ising-like screw chain antiferromagnet $SrCo_2V_2O_8$ in a transverse magnetic field, Phys. Rev. Lett. **123**, 067203 (2019).

[21] A. Okutani, H. Onishi, S. Kimura, T. Takeuchi, T. Kida, M. Mori, A. Miyake, M. Tokunaga, K. Kindo, and M. Hagiwara, Spin excitations of the $S = 1/2$ one-dimensional Ising-like antiferromagnet $BaCo_2V_2O_8$ in transverse magnetic fields, J. Phys. Soc. Jpn. **90**, 044704 (2021).

[22] Z. Wang, C.-M. Halati, J.-S. Bernier, A. Ponomaryov, D. I. Gorbunov, S. Niesen, O. Breunig, J. M. Klopf, S. Zvyagin, T. Lorenz, A. Loidl, and C. Kollath, Experimental observation of repulsively bound magnons, Nature **631**, 760 (2024).

[23] R. Kuhn, Über Verdazyle und verwandte Stickstoffradikale, Angew. Chem. **76**, 691 (1964).

[24] M. Shoji et al., A general algorithm for calculation of Heisenberg exchange integrals J in multispin systems, Chem. Phys. Lett. **432**, 343-347 (2006).

[25] H. Yamaguchi, T. Okubo, K. Iwase, T. Ono, Y. Kono, S. Kittaka, T. Sakakibara, A. Matsuo, K. Kindo, and Y. Hosokoshi, Various regimes of quantum behavior in an $S=1/2$ Heisenberg antiferromagnetic chain with fourfold periodicity, Phys. Rev. B **88**, 174410 (2013).

[26] H. Yamaguchi, H. Miyagai, T. Shimokawa, K. Iwase, T. Ono, Y. Kono, N. Kase, K. Araki, S. Kittaka, T. Sakakibara, T. Kawakami, K. Okunishi, and Y. Hosokoshi, Field-induced incommensurate phase in the strong-rung spin ladder with ferromagnetic legs, Phys. Rev. B **89**, 220402(R) (2014).

[27] H. Yamaguchi, S. C. Furuya, S. Morota, S. Shimono, T. Kawakami, Y. Kusanose, Y. Shimura, K. Nakano, and Y. Hosokoshi, Observation of thermodynamics originating from a mixed-spin ferromagnetic chain, Phys. Rev. B **106**, L100404 (2022).

[28] H. Tsukiyama, S. Morota, S. Shimono, Y. Iwasaki, M. Hagiwara, Y. Hosokoshi, and H. Yamaguchi, Crystal structures and magnetic properties of verdazyl-based complexes with transition metals, Phys. Rev. Mater. **6**, 094417 (2022).

[29] A. Abragam and M. H. L. Pryce, The theory of paramagnetic resonance in hydrate cobalt salts, Proc. R. Soc. A **206**, 173 (1951).

[30] M. E. Lines, Magnetic properties of $CoCl_2$ and $NiCl_2$, Phys. Rev. **131**, 546 (1963).

[31] Y. Kojima, M. Watanabe, N. Kurita, H. Tanaka, A. Matsuo, K. Kindo, and M. Avdeev, Quantum magnetic properties of the spin-1/2 triangular-lattice antiferromagnet $Ba_2La_2CoTe_2O_{12}$, Phys. Rev. B **98**, 174406 (2018).

[32] X. C. Liu, Z. W. Ouyang, T. T. Xiao, J. J. Cao, Z. X. Wang, Z. C. Xia, Z. Z. He, and W. Tong, Magnetism and ESR of the $S_{eff}= 1/2$ antiferromagnet $BaCo_2(SeO_3)_3·3H_2O$ with dimer-chain structure, Phys. Rev. B **105**, 134417 (2022).

[33] S. Morota, Y. Iwasaki, M. Hagiwara, Y. Hosokoshi, and H. Yamaguchi, Magnetic anisotropy in a verdazyl-based complex with cobalt(II), J. Phys. Soc. Jpn. **92**, 054705 (2023).

[34] H. Yamaguchi, Y. Tominaga, A. Matsuo, S. Morota, Y. Hosokoshi, M. Hagiwara, and K. Kindo, Ladder-based two-dimensional spin model in a radical-Co complex,

Phys. Rev. B **107**, 174422 (2023).
[35] H. Yamaguchi, Y. Tominaga, T. Kida, K. Araki, T. Kawakami, Y. Iwasaki, K. Kimura, and M. Hagiwara, Realization of a spin-1/2 Kondo necklace model with magnetic field-induced coupling switch, Phys. Rev. Res. **7**, L012023 (2025).
[36] M. Ito, M. Mito, H. Deguchi, and K. Takeda, The numerical comparison of magnetic susceptibility and heat capacity of TMNIN with the result of a quantum Monte Carlo method for the Haldane system, J. Phys. Soc. Jpn. **63**, 1123 (1994).
[37] M. Orendáč, R. Tarasenko, V. Tkáč, A. Orendáčová, and V. Sechovský, Specific heat study of the magnetocaloric effect in the Haldane-gap $S = 1$ spin-chain material $[Ni(C_2H_8N_2)_2NO_2](BF_4)$, Phys. Rev. B **96**, 094425 (2017).
[38] N. Ishimura, H. Shiba, Dynamical correlation functions of one-dimensional anisotropic Heisenberg model with spin 1/2. I: Ising-like antiferromagnets, Prog. Theor. Phys. **63**, 743 (1980).
[39] H. J. Schulz, Dynamics of Coupled Quantum Spin Chains, Phys. Rev. Lett. **77**, 2790 (1996).